\documentclass[aps, prb, twocolumn,groupedaddress]{revtex4}
\usepackage{graphicx}
\usepackage{amsmath}
\usepackage{amssymb}

\newcommand{\bb}{\begin{equation}}
\newcommand{\ee}{\end{equation}}
\newcommand{\ba}{\begin{eqnarray*}}
\newcommand{\ea}{\end{eqnarray*}}

\usepackage{graphicx}% Include figure files
\usepackage{dcolumn}% Align table columns on decimal point
\usepackage{bm}% bold math
\usepackage{longtable}

\begin{document}

%\preprint{}

%\title{The simulation of the surface tension of spherical drops from test ellipsoidal deformations}

\title{Evidence for the role of fluctuations in the thermodynamics of nanoscale drops and the implications in computations of the surface tension}

\author{Jos\'e G. \surname{Sampayo}$^\dag$}
\author{Alexandr \surname{Malijevsk\'y}$^{\dag,\ddag}$}
\author{Erich A. \surname{M\"{u}ller}$^\dag$}
\author{Enrique \surname{de Miguel}$^\S$}
\author{George \surname{Jackson$^\dag$}\footnote{Corresponding author: g.jackson@imperial.ac.uk}}

\affiliation{$^\dag$Department of Chemical Engineering, Imperial College London,
South Kensington, London SW7 2AZ, UK\\
$^\S$Departamento de F\'{\i}sica Aplicada, Facultad de Ciencias
Experimentales, Universidad de Huelva, 21071 Huelva, Spain\\
$^\ddag$E. H{\'a}la Laboratory of Thermodynamics, Institute of Chemical Process Fundamentals of ASCR, 16502 Prague 6, Czech Republic}


%\date{9 November 2007}% It is always \today, today,
\begin{abstract}

Test area deformations are used to analyse vapour-liquid interfaces of Lennard-Jones particles by molecular dynamics simulation. For planar vapour-liquid interfaces the change in free energy is captured by the average of the corresponding change in energy, the leading-order contribution. This is  consistent with the commonly used mechanical (pressure tensor) route for the surface tension. By contrast for liquid drops one finds a large second-order contribution associated with fluctuations in energy. Both the first- and second-order terms make comparable contributions, invalidating the mechanical relation for the surface tension of small drops. The latter is seen to  increase above the planar value for drop radii of $\sim 8$ particle diameters, followed by an apparent weak maximum and slow decay to the planar limit, consistent with a small negative Tolman length.

%Valid PACS numbers may be entered using the \verb+\pacs{#1}+ command.
\end{abstract}

\pacs{Valid PACS appear here}% PACS, the Physics and Astronomy
\keywords{Density functional theory, fundamental measure theory, Monte Carlo and molecular dynamics simulation, Lennard-Jones fluid, interfaces, vapour-liquid equilibria, drops, nucleation, finite-size effects, Laplace relation, Kelvin equation, Tolman length.}
%Use showkeys class option if keyword

%\maketitle must follow title, authors, abstract, \pacs, and %\keywords
\maketitle

%\setion{}

It is striking that though almost a century has passed since Gibbs formulated his thermodynamic theory of curved interfaces, there is still widespread controversy about the dependence of the surface tension on the curvature (size of a drop), and the validity of the mechanical route to the surface tension~\cite{RowlinsonWidom1982, SchofieldHenderson1982, Croxton}.
The formal approach of Gibbs is intimately connected with the relations of Laplace, $\Delta p = 2\gamma_s/R_s$, and Tolman,  $\gamma(R)/\gamma_\infty=1-2\delta_\infty/R+\cdots$, for drops of radius $R$.
Here, $\Delta p=p_l-p_g$ is the pressure difference inside (l) and outside (g) the drop, $\gamma_s=\gamma(R_s)$ is the interfacial tension associated with the surface of tension $R_s$, $\gamma_\infty$ the  value for the planar gas-liquid surface, and the Tolman~\cite{Tolman1949} length $\delta_\infty$ is defined relative to the radius of the equimolar surface $R_e$ as $\delta_\infty=\lim_{R_s\rightarrow\infty}(R_e-R_s)$.

There are 3 basic routes to the definition of the tension~\cite{RowlinsonWidom1982}:
thermodynamic (Gibbs and Tolman), mechanical (Laplace and Young), and statistical mechanical (density functional and related theories). The thermodynamic and mechanical routes are macroscopic theories so there has been much debate about their applicability to small systems such as nanoscale liquid drops or bubbles. One key question is whether the mechanical relations based on the pressure virial (formulated in terms of the appropriate tensorial components) that make use of the concept of the bulk pressure of the coexisting states are appropriate at these length scales for curved surfaces.

%It is well known that in the thermodynamic limit a drop is unstable with respect to the uniform state. The equilibrium of the system liquid drop-vapour corresponds to the unstable size of the drop ($\delta^2\Omega/\delta R^<0$), where $\Omega$ is the grand potential of the system) that can be then identified as a critical nucleus. In order to stabilise the drop in an open system ($\mu VT$ ensemble) it is needed to turn on a radial external field. On the other hand, in a finite system ($NVT$ ensemble) the free energy posses a minimum the location of which highly depends on a system size.

While the Laplace equation essentially defines the ratio $\gamma_s/R_s$, the first-order form of Tolman's theory is appropriate only for sufficiently large drops. One can view $\delta_\infty$ as the leading-order correction to the tension of a planar surface. Despite its fundamental role in studies of interfacial properties of curved surfaces and  theories of nucleation, there is still much controversy as to even  the \emph{sign} of $\delta_\infty$.
Microscopic statistical mechanical approaches including
square gradient theories (SGTs)~\cite{Falls1981,Guermeur1985}, curvature
expansions of the planar interface~\cite{Blokhuis1992a,Blokhuis1992b}, and density functional theories (DFTs), including local~\cite{Lee1986,Oxtoby1988,Koga1998} and non-local~\cite{Bykov2002,Wu2008,Malijevsky2009} treatments, have led to conflicting views on the magnitude and sign of $\delta_\infty$, as well as the curvature dependence of the surface tension. The widely accepted view from this body of work is that $\delta_\infty \lesssim 0$ and that there is a small maximum in $\gamma_s(R)$ as the drop radius is decreased, then followed by a sharp decrease. This is supported by studies on the penetrable sphere model~\cite{Hemingway1981} (which can be solved exactly at the mean-field level at zero temperature) where one finds a negative Tolman length ($\delta_\infty=-\sigma/2$), with $\sigma$ the molecular diameter.

By contrast, the vast majority of computer simulation studies suggest that $\delta_\infty>0$. In most simulations of liquid drops, the mechanical route to the interfacial tension is employed, usually involving an integration of the gradient of the normal component of the pressure tensor from the centre of the drop to the bulk vapour phase~\cite{Thompson1984,Nijmeijer1992,Lei2005, Vrabec2006}. In this case one predicts a monotonous decrease in the surface tension with increasing curvature (decreasing drop radius) from the planar limit (infinite radius); this would correspond to $\delta_\infty>0$ throughout.
As was pointed out early on by Schofield and Henderson~\cite{SchofieldHenderson1982}
there are fundamental problems in employing local pressure tensors and the associated definition of the internal pressure for microscopic (high curvature) drops. This leads to a mismatch in the free energy of the formation of a drop determined via the mechanical and thermodynamic routes as observed in simulation~\cite{Wolde1998}.
Macroscopic thermodynamic routes based on a combination of the Laplace and Tolman relations have been employed~\cite{Thompson1984} but also suffer from the ill-definition of the internal pressure and density of the liquid. One can also estimate the interfacial tension from the free energy change accompanying a volume deformation of spherical surfaces using a virial-like expression~\cite{Bardouni2000}; these results for the surface tension are in disagreement with those obtained from the direct mechanical route. Recent grand canonical simulations~\cite{Binder2009}, and a thermodynamic analysis of large drops based on Laplace-Tolman relations~\cite{Giessen2009} both now appear to suggest a small negative $\delta_\infty$, which is consistent with the findings of DFT.

The aim of this Letter is to use a new method for the calculation of the surface tension of small liquid drops in molecular simulation, highlighting the role played by the fluctuations in the energy of deformation. The method relies on the thermodynamic definition of the surface tension and is thus free from the inconsistencies associated with the application of the mechanical route. A variant of the test-area (TA) method~\cite{Gloor2005} is used where small virtual perturbations are made in the box dimensions of systems with interfaces to obtain the change in free energy associated with the corresponding change in surface area. For a fluid drop of radius $R$ the change in the Helmholtz free energy $F$ is expressed thermodynamically as~\cite{RowlinsonWidom1982}
\bb
dF = -SdT - p_gdV_g + p_ldV_l + \mu dN + \gamma dA + C dR \, ,
\ee
\noindent with $S$ the entropy, $V_{g,l}$ the vapour and liquid volumes, $T$ the temperature,  $\mu$ the chemical potential, $N$ the number of particles, $A$ the interface area, and $C$ the conjugate variable for $R$. The surface tension of a drop is given by
\bb
{\left( \frac{\partial F}{\partial A}\right)}_{NVT} = \gamma_s \, ,
\ee
where the minimal interfacial tension $\gamma_s$ defines  $R_s$ and corresponds to taking $C=0$. The change in free energy $\Delta F$ due to a virtual change in area $\Delta A$ can be expressed as the average of the Boltzmann factor of the corresponding change in configurational energy $\Delta U$~\cite{Gloor2005}
\begin{eqnarray}
\Delta F &=& - kT \ln {\left\langle \exp {\left( -\frac{\Delta U}{kT} \right)} \right\rangle} \label{Boltz}\\
&=& \label{Boltz2} {\left\langle \Delta U \right\rangle} - \frac{1}{2kT} {\left\{ {\left\langle \Delta U^2 \right\rangle} - {\left\langle \Delta U \right\rangle}^2 \right\}} \\
&+& \frac{1}{6(kT)^2} {\left\{ {\left\langle \Delta U^3 \right\rangle} - 3 {\left\langle \Delta U^2 \right\rangle} {\left\langle \Delta U \right\rangle}  +2 {\left\langle \Delta U \right\rangle}^3 \right\}}\,. \nonumber
\end{eqnarray}
The averages are over configurations of the unperturbed reference system. In Eq. (\ref{Boltz2}) $\Delta F$ is expressed as a perturbation series to ${\cal O} \left( \left\langle \Delta U^3 \right\rangle \right)$, where the first-order average of the change in energy is $\Delta F_1={\left\langle \Delta U \right\rangle}$, the second-order energy fluctuation term is $\Delta F_2=-{\left\{ {\left\langle \Delta U^2 \right\rangle}- {\left\langle \Delta U \right\rangle}^2 \right\}}/(2kT)$, and the third-order contribution is denoted by $\Delta F_3$. The full Boltzmann form, Eq.~(\ref{Boltz}), is employed in, e.g., the test-particle approach for the chemical potential~\cite{Widom1963}, or the volume perturbation method for the pressure~\cite{Eppenga1994} and the pressure tensor~\cite{deMiguel2006}.
\begin{figure}
  \includegraphics[angle=0, width={0.35\textwidth}]{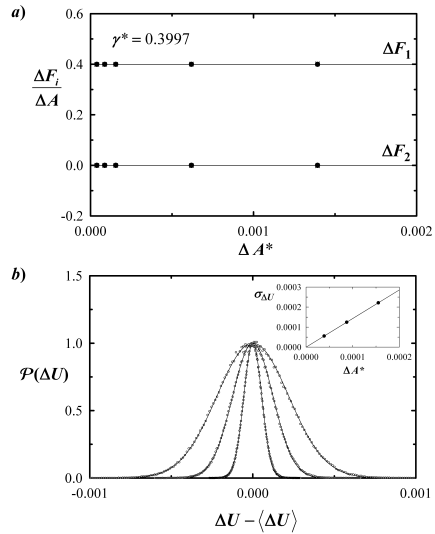}\\
{\caption [TA deformations for planar interface of LJ particles]
{\label{Figure1} Test-area deformations of a planar liquid-vapour interface of the LJ-TS fluid. MD simulations of $N = 749$ particles in  a periodic box of dimensions
$L_x = L_y = 7.885\sigma$ and $Lz = 6L_x$ at $T^*=kT/\epsilon=0.8$ over $3 \times 10^{6}$ timesteps. The deformations correspond to changes in the box dimensions (particle coordinates) of $L_x'=L_x\sqrt{1+\Delta A^*}$, $L_y'=L_y\sqrt{1+\Delta A^*}$, and $L_z'=L_z/(1+\Delta A^*)$.
a) The contributions $\Delta F_1/\Delta A$ and $\Delta F_2/\Delta A$ to the change in free energy per unit area (in units of $\epsilon/\sigma^2)$ ($\Delta A^*<0$, $+$; $\Delta A^*>0$, $\times$; average, $\bullet$). The interfacial tension $\gamma^*=\gamma \sigma^2/\epsilon$ is obtained by extrapolation to $\Delta A^*=0$. b) The distribution ${\cal P} {\left( \Delta U \right)}$ of the change in energy (relative to its average in units of $\epsilon$) scaled at the maximum peak height for different relative deformations $\Delta A^*$. The width (standard deviation, $\sigma_{\Delta U}$) is depicted in the inset.}}
\end{figure}

The tension is obtained as the
change in free energy per unit area for infinitesimal
perturbations to ${\cal O} \left( \left\langle \Delta U^3 \right\rangle \right)$:
\begin{eqnarray}
\gamma = \lim_{\Delta A \rightarrow 0} \frac{\Delta F}{\Delta A}
= \lim_{\Delta A \rightarrow 0} \left\{ \frac{\Delta F_1}{\Delta A}+\frac{\Delta F_2}{\Delta A}+\frac{\Delta F_3}{\Delta A} \right\} \label{Pert} \,.
\end{eqnarray}
The leading term, $\Delta F_1={\left\langle \Delta U \right\rangle}$, corresponds to the mechanical work involved in changing the area of the interface, which can be directly associated with the so-called virial expression for the tension~\cite{Lekner1977} (expressed in terms of averages of the appropriate components $\alpha$ of the virial, ${\left\langle x_\alpha \left( {\rm d} U / {\rm d} x_\alpha \right) \right\rangle}$, at the Hookian linear-response level). The corresponding entropic contribution due to the deformation is~\cite{Lekner1977}:
$T \Delta S= \left\{ \left\langle U \right\rangle \left\langle \Delta U \right\rangle - \left\langle U \Delta U \right\rangle \right\}/(kT)$.

In the case of the a planar interface, it is well known that
the interfacial tension can be obtained formally from the virial expression~\cite{RowlinsonWidom1982,Lekner1977}, i.e., entirely from the leading-order contribution of Eq.~(\ref{Pert}). This is exemplified for a planar vapour-liquid interface of Lennard-Jones particles (of diameter $\sigma$ and well depth $\epsilon$, truncated and shifted TS at $r_c=2.5\sigma$) as shown in Fig. 1. A planar interface is first stabilised during an $NVT$ molecular dynamics (MD) simulation of the inhomogeneous system with a liquid slab in the centre of the box separated by two vapour regions. The change in configurational energy due to small test changes in the dimensions of the box such that the area is increased or decreased at fixed overall volume is then computed to estimate the various contributions in Eq.~(\ref{Pert}); the limit of infinitesimal deformations is obtained by extrapolation to $\Delta A \rightarrow 0$. From Fig. 1a it is clear that only the leading {\it mechanical} term in $\Delta F_1 / \Delta A$ contributes to the interfacial tension of a planar interface, confirming the validity of the pressure-tensor route in this case. The fluctuation term $\Delta F_2 / \Delta A$ is very small by comparison and does not contribute to the tension in an appreciable way; this is also true for the third-order term.
In Fig. 1b we plot the distribution ${\cal P} {\left( \Delta U \right)}$ of the change in configurational energy (relative to ${\left\langle \Delta U \right\rangle}$) for different area perturbations; the distribution is well represented by a Gaussian, the width of which ($\sigma_{\Delta U}$) decreases to zero with $\Delta A \rightarrow 0$, consistent with a very small $\Delta F_2 / \Delta A \sim 1 \times 10^{-6}\epsilon / \sigma^2$.

The overall physical picture is fundamentally different for a nanosized spherical drop of liquid in contact with its vapour. Once the drop has been stabilised its size can be characterised from the density profile $\rho(r)$ as a function of the distance $r$ from its centre by calculating the Gibbs dividing surface $R_e^3=(\rho_v - \rho_l)^{-1} \int {\rm d} r \, r^3 {\rm d} \rho(r) / {\rm d} r$, corresponding to an area of $A=4\pi R_e^2$. Virtual perturbations from the equilibrium spherical drop geometry are made with test changes in the dimensions of the simulation cube:  two of the Cartesian axes are decreased (or increased) in length and the third is increased (or decreased) such that the overall volume remains constant. The perturbed states correspond to ellipsoidal drops of prolate (or oblate) shape which always have larger surface areas than the original drop, $\Delta A>0$. This essentially corresponds to the longest $P_2$ (Legendre polynomial) capillary-wave oscillations possible for the drop~\cite{Croxton}; the capillary-wave surface tension is equivalent to the thermodynamic one at least to leading order in curvature ${\cal O}(1/R)$. Averages are then accumulated over very long runs of $\sim 1.5 \times 10^{9}$ timesteps, corresponding to microsecond runs for typical molecular parameters. The term $\Delta F_1 / \Delta A$ is more than two orders of magnitude larger in the case of the drop than for a planar interface systems of comparable size (cf. Fig. 1a and Fig. 2a). The most significant difference is the large contribution from the second-order energy ``fluctuation" term $\Delta F_2 / \Delta A$ for the drop, which was negligible for the planar interface; this term is now comparable in magnitude but of opposite sign to the first-order term. The third-order terms remains essentially negligible. Thus both the first- and second-order terms  contribute to the surface tension of the drop. A thermodynamic characteristic of the drop is thus the non-vanishing (and large) fluctuation term, which is clearly an indication of an additional entropic contribution. This can be seen in the distribution of the change in configurational energy for different TA perturbations (Fig. 2b). The data are again well described as Gaussians, but now the width does not vanish in the limit of infinitesimal deformations, i.e., $\lim_{\Delta A \rightarrow 0} \sigma_{\Delta U} \ne 0$ and correspondingly $\lim_{\Delta A \rightarrow 0} \Delta F_2 \ne 0$. The fact that $\lim_{\Delta A \rightarrow 0} \Delta F_3 \sim 0$ for both the planar and curved systems suggests symmetrical Gaussians.

\begin{figure}
  \includegraphics[angle=0, width={0.35\textwidth}]{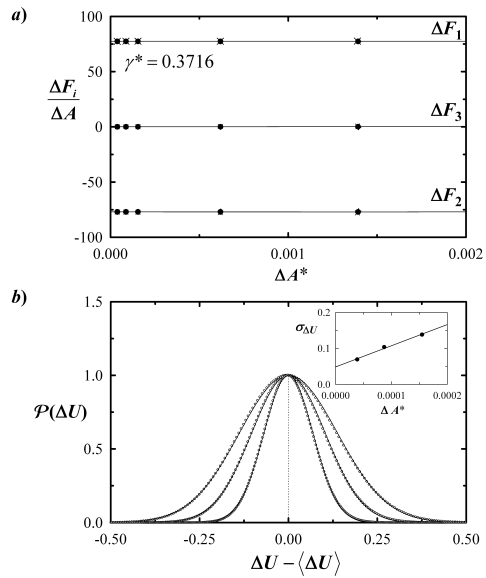}\\
{\caption [TA deformations of a spherical drop of LJ particles]
{\label{Figure2}
Test-area ellipsoidal deformations of a spherical drop of LJ-TS liquid of radius $\left\langle R_e \right\rangle = 5.55\sigma$ in coexistence with its vapour. MD simulations of $N = 749$ particles in a periodic box of dimensions
$L_x = L_y = L_z=20\sigma$ at $T^*=0.8$ over $1.5 \times 10^{9}$ timesteps. The deformations correspond to changes in the box dimensions (particle coordinates) of $L_i'=L_i\sqrt{1+\Delta A^*}$, $L_j'=L_j\sqrt{1+\Delta A^*}$, and $L_k'=L_k/(1+\Delta A^*)$ (where $i$, $j$, $k$ denotes any of the Cartesian axes).
a) The contributions $\Delta F_1/\Delta A$, $\Delta F_2/\Delta A$, and $\Delta F_3/\Delta A$ to the change in free energy per unit area (in units of $\epsilon/\sigma^2)$ (prolate, $+$; oblate, $\times$; average, $\bullet$). The tension $\gamma^*=\gamma \sigma^2/\epsilon$ is obtained by extrapolation to $\Delta A^*=0$. b) The distribution ${\cal P} {\left( \Delta U \right)}$ of the change in energy scaled at the maximum peak height for different relative deformations $\Delta A^*$; the width $\sigma_{\Delta U}$ is depicted in the inset. The Gaussians for the planar interface are shown dotted (note the very small scale in comparison).}}
\end{figure}

\begin{figure}
  \includegraphics[angle=0, width={0.4\textwidth}]{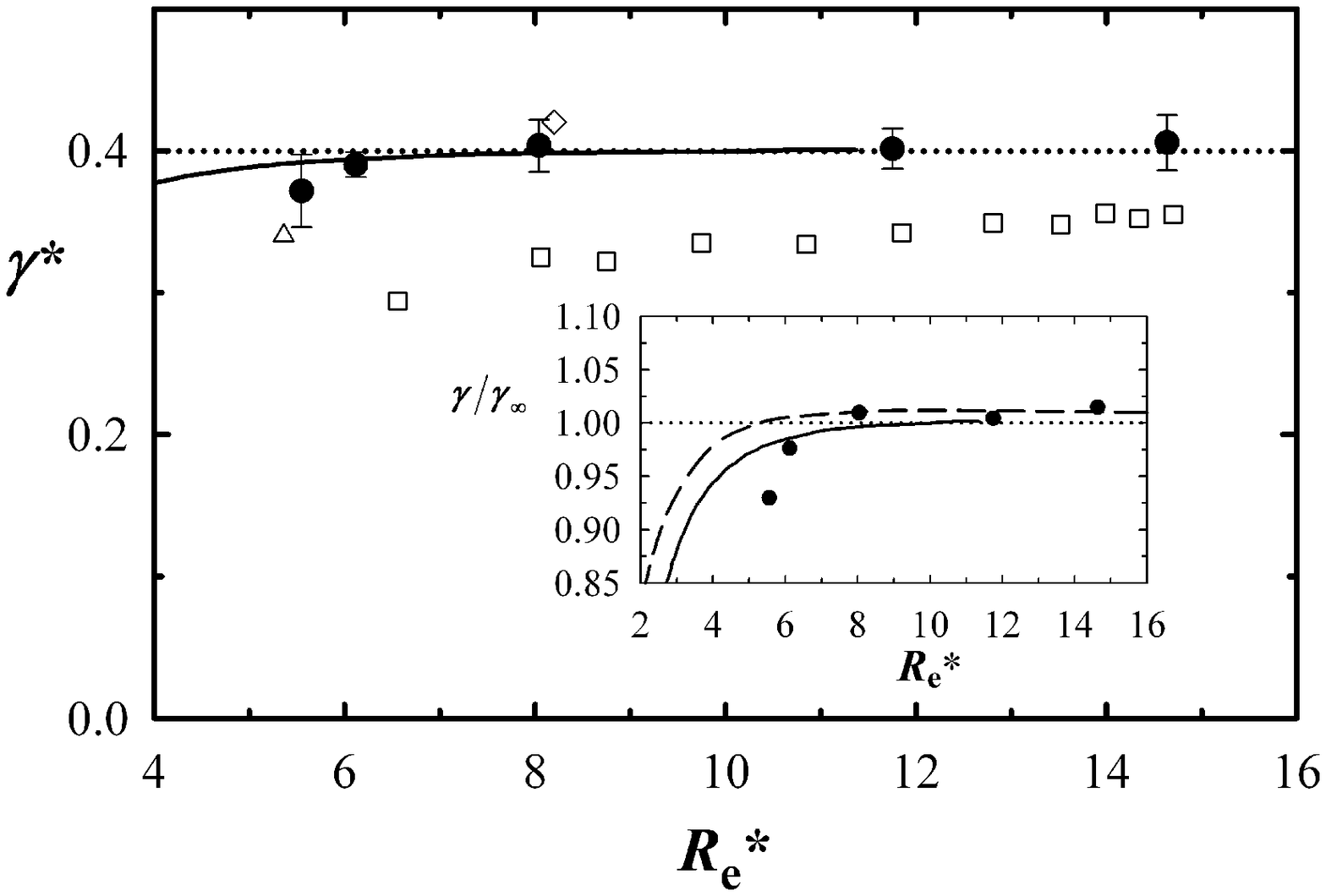}\\
{\caption [TA surface tension of drops]
{\label{Figure3}
The surface tension of spherical drops of LJ-STS fluids with average radii $\left\langle R_e/\sigma \right\rangle = 5.55$, 6.12, 8.04, 11.7 and 14.6 at $T^*=0.8$ from TA ellipsoidal deformations ($\bullet$), compared with the values from the mechanical route~\cite{Vrabec2006} ($\square$), and the data of Thomson~{\it et al.}~\cite{Thompson1984} ($\vartriangle$),  El~Bardouni~{\it et al.}~\cite{Bardouni2000} ($\lozenge$), and Schrader {\it et al.}~\cite{Binder2009} (continuous curve); the planar limit is shown dotted. The predictions of FMT~\cite{Malijevsky2009} are depicted in the inset (dashed curve).}}
\end{figure}

The dependence of the surface tension computed from  ellipsoidal deformations as a function of the drop size (for systems with $N=749$ to 11,334) is depicted in Fig. 3. Here the tension is computed for $R_e$ rather than $R_s$, though $\gamma_e = \gamma_s$ to ${\cal O}(1/R^2)$. The behaviour obtained with our thermodynamic TA approach does not support the findings obtained from a standard pressure-tensor route (e.g., the recent MD data of Vrabec {\it et al.}~\cite{Vrabec2006}). This is in line with the concerns of Schofield and Henderson~\cite{SchofieldHenderson1982}, and others~\cite{Blokhuis1992b,Wolde1998,Reiss2004} about the inadequacy of the mechanical route for very small systems. For drops larger than $R_e \sim 8$ we observe values of the surface tension which appear to be slightly larger than the planar limit, $\gamma(R)>\gamma_\infty$; because the tension has to converge to $\gamma_\infty$ when $R_e \rightarrow \infty$ this suggests a non-monotonic behaviour of the tension with increasing curvature, and a corresponding weak maximum. Our values are consistent with the data point reported by El Bardouni {\it et al.}~\cite{Bardouni2000} estimated from the surface free energy change, and with the small maximum observed by Schrader~{\it et al.}~\cite{Binder2009} using a Landau free energy approach in the canonical ensemble (though the authors do not comment explicitly on this point). The calculations of the
tension of curved interfaces from curvature corrections, SGT and DFT (which have been brought into question because of their failure to reproduce existing simulation data) are now supported by our data. In the inset of Fig. 3 we compare the TA data for the surface tension of drops with those from a non-local DFT using fundamental measure theory (FMT)~\cite{Malijevsky2009}; a maximum is predicted with FMT at $R_e \sim 10$.

Three main conclusions can be gleaned from our study. Firstly, there is clearly a large fluctuation contribution to the interfacial tension of nanoscale spherical drops (and most likely other curved surfaces) in addition to the underlying first-order (mechanical) contribution which fully describes the planar interface. Such contributions from fluctuations in the energy are not found in the planar limit to any significant degree. As a consequence, our results do not support the validity of a mechanical (pressure-tensor) route to the interfacial tension for surfaces of high curvature such as small drops. This is in line with the warning of Blokhuis and Bedeaux~\cite{Blokhuis1992b} that the use of a mechanical approach in this context ``is still a matter of concern" and that it is ``advisable not to use the pressure-tensor whenever this can be avoided". Our data are not consistent with the monotonic dependence of the surface tension with curvature obtained from a mechanical treatment. As well as contributions in  ${\left\langle x_\alpha \left( {\rm d} U / {\rm d} x_\alpha \right) \right\rangle}$, the correct ``virial" expression for the surface tension would have to contain terms in averages of the type
${\left\langle x_\alpha \left( {\rm d} U / {\rm d} x_\alpha \right) \right\rangle}{\left\langle x_\beta \left( {\rm d} U / {\rm d} x_\beta \right) \right\rangle}$ and ${\left\langle x_\alpha x_\beta \left( {\rm d} U / {\rm d} x_\alpha \right) \left( {\rm d} U / {\rm d} x_\beta \right) \right\rangle}$, which would involve up to four-body correlations for pair-wise additive potentials. This suggests that there are additional contributions to the change in the entropy due to the deformation of small drops involving quadratic terms in $\Delta U$:
${\left\langle \Delta U^2 \right\rangle}$, ${\left\langle \Delta U \right\rangle^2}$, ${\left\langle U \Delta U^2 \right\rangle}$, ${\left\langle \Delta U \right\rangle} {\left\langle U \Delta U \right\rangle}$, ${\left\langle U \right\rangle} {\left\langle \Delta U^2 \right\rangle}$, and ${\left\langle U \right\rangle} {\left\langle \Delta U \right\rangle^2}$.
As a final point, the rise of the surface tension above that of the planar limit after a certain drop size would be consistent with a negative Tolman length. Our data for the larger drops suggests a value of $\delta_\infty\sim -0.2 \pm 0.3$. Though the statistical uncertainty is large, our finding supports the exact mean-field predictions for the penetrable sphere model~\cite{Hemingway1981}, and is consistent with the latest accurate value of $-0.10 \pm 0.02$ determined from the Laplace relation for a large $N=100,000$ particle system~\cite{Giessen2009}.

%\begin{acknowledgments}
\noindent We are very appreciative to Jim Henderson for useful discussions. Financial support from the Czech, Mexican, Spanish and UK councils is gratefully acknowledged.
%\end{acknowledgments}

%% Create the reference section using BibTeX:
%\bibliography{bibexperimental}

\bibliographystyle{apsrev}

\end{document}